\documentclass[letterpaper,11pt]{article}
\usepackage{jheppub_mod}
\usepackage{graphicx, amsmath, amssymb, amstext}
\usepackage{relsize, float, multirow}
\usepackage{array, epstopdf, hyperref, }
\usepackage{caption, subcaption, mathtools}
\usepackage{newtxmath}

\newcommand{\kB}{k_{\rm B}}

\title{New CMB spectral distortion constraints on decaying dark matter
  with full evolution of electromagnetic cascades before recombination}
\author[a]{Sandeep Kumar Acharya,} 
\author[a]{Rishi Khatri}
\affiliation[a]{Department of Theoretical Physics, Tata Institute of 
Fundamental Research, Mumbai 400005, India}
\emailAdd{sandeepkumar@theory.tifr.res.in, khatri@theory.tifr.res.in}
\date{\today}

\date{\today}

\abstract{
	Current constraints on  energy injection in the form of energetic
        particles before the epoch of recombination using CMB spectral
        distortions  assume that all energy goes into $y$ and $\mu$-type
        distortions. We revisit these constraints with exact calculations
        of the spectral distortions by evolving the electromagnetic 
        cascades. The actual spectral distortion differs in shape and
        amplitude from the $y$-type distortion and depends on the energy
        and nature of injected particles. The constraints on the energy
        injection processes such as dark matter decay can be relaxed by as
        much as a factor of 5.}

\begin{document}                         
\maketitle


\section{Introduction.}
  In the standard $\Lambda$CDM cosmology, the cosmic microwave background
(CMB) has an almost perfect blackbody spectrum, a consequence of almost complete thermal equilibrium in the early
Universe. Any deviation from equilibrium in the CMB spectrum is exponentially suppressed at redshifts $z\gtrsim 2\times 10^6$
\cite{Sz19701,dd1982,Chluba:2011hw,ks2012}. The Planck spectrum once
created is preserved by the expansion of the Universe. At redshifts
$z\lesssim 2\times 10^6$, it is no longer possible to restore the
equilibrium disturbed by, for example, injection of energetic standard model
particles such as photons, electrons and positrons. It therefore becomes possible
to create deviations in the CMB from a Planck spectrum. By observing these
spectral distortions of the CMB, we can learn about the
physical processes which injected energy into the CMB at $z\lesssim 2\times
10^6$. The
high entropy per baryon in our Universe, quantified by the photon-to-baryon
number density ratio ($\sim 10^9$), means that the recombination dynamics is
controlled by the energetic photons in the Wien tail of the Planck
spectrum with the recombination starting only when the temperature of the
CMB is more than a factor of 30 smaller than the Lyman-$\alpha$ energy
\cite{Zks1969,Peebles1969}. As
a result, the recombining and subsequent Universe, with most neutral atoms
in the ground state, is optically thin to the
bulk of the photons in the CMB since they are at energies much smaller compared
to the energy of the 
first excited state of neutral atoms. The CMB spectral distortions, and the
information about the energy injection processes in the early Universe, are therefore preserved
through recombination and are observable today. 

 The measurement of the CMB monopole spectrum in 1990s by the Far Infrared Absolute Spectrometer (FIRAS) onboard Cosmic Background
 Explorer (COBE) satellite \cite{Cobe1994,F1996} remains the best constraint on the deviation
 of the CMB monopole spectrum from a blackbody until today. The COBE-FIRAS data has
 been used by numerous authors to constrain many exotic as well as
 standard model
 energy injection processes in the early Universe (e.g. \cite{Hs1993,FRT20031,DKC2016}, see
 \cite{Sunyaev:2013aoa} for a review). Almost all calculations until now
 have assumed that the injected energy ends up heating the electrons \citep{Sz1969,Sz19701,Is19752,Bdd1991,Chluba:2011hw,Ks2012b,Chluba:2013vsa}
 irrespective of the initial energy or nature of
 the injected particles. The Compton scattering of CMB photons on heated electrons
 produces $y$-type distortions which then thermalize to $i$-type and
 $\mu$-type (Bose-Einstein spectrum) distortions \cite{Ks2012b}. We will
 call $y$, $i$ and $\mu$-type distortions
 collectively as $yim$-distortions and the assumption that all energy goes
 into heat or $yim$-distortions as the $yim$-approximation. The amplitude of the $yim$-distortions
 depends on the amount of energy injected while the shape depends on the
 redshift of energy injection, with departure from the $y$-type distortion
 becoming important at $z\gtrsim 10^4$. The information hidden
   beyond $y$ and $\mu$-type distortions, in the $i$-type distortions, has
   been explored previously by \cite{Ks2012b,Ks2013,Chluba:2013vsa,Cj2014}.

 We have recently shown that this simple picture of $yim$-distortions is
   incomplete  \cite{AK2018}. In particular, when the energy is injected in
  the form of standard model particles, with energy much greater than the average
  energy of the CMB photons, not all of the energy goes into the $yim$-distortions as is usually assumed. A significant fraction of energy
  is lost by the particles as they collide with the background electrons,
  ions and photons and make their way down in energy towards thermalization
  with the background. The Compton scattering of relativistic electrons,
  produced in the cascade, with
  the CMB results in spectral distortions with distinctly different shape and amplitude
  compared to the $yim$-distortions. We call these distortions  non-thermal
  relativistic or $ntr$-type spectral distortions. The
  shape and amplitude of the $ntr$-type spectral distortions depend on the
  energy, type  and redshift of injected particles \cite{AK2018}. In
  particular, a fraction of the energy goes into the high frequency Wien tail of
  the spectral distortions resulting in decrease in the amplitude of the
  distortions in the main observable CMB frequency range compared to the
  $yim$-approximation.  We
  therefore expect significant corrections to the existing constraints on
  energy injection scenarios in the early Universe. 

A particularly
  interesting scenario which injects energetic particles is decay of
  unstable dark matter particles.    The dark matter direct
  detection experiments, looking for Weakly Interacting Massive Particles
  (WIMPS)  have so far not yielded any detection and the allowed parameter space
  for WIMPs is gradually being ruled out.  This has
  motivated physicists to look beyond the simplest thermal WIMP models. One
  way around the direct detection constraints is
  Super-WIMPs (SWIMP), where instead of producing the usual lightest stable
  particle in the theory in a beyond standard model theory like supersymmetry or
  Kaluza-Klein models, the thermally produced WIMP is the  next-to-lightest
  particle which then decays to the lightest stable particle
  of the theory with the right abundance needed for dark matter \cite{FRT2003,FRT20031}. For a review of dark matter models, see \cite{F2010}. 
 We
  can escape the direct detection constraints if the lightest particle, which would be the
  dark matter today, has much weaker interactions with the standard model
  particles i.e. is a SWIMP. We could in general have additional unstable particles which were
  produced in the early Universe, with lifetime of few years to hundreds of
  thousands of years, which would decay not into SWIMPs + standard model
  particles but completely into standard model particles. More generally,
  we would like to constrain energy injection in the form of different
  standard model particles  over the entire history of
  the Universe and look for deviations from standard cosmology as a probe
  of beyond standard model physics.  Another interesting example of energy
  injection in the form of energetic photons and other particles is from evaporating primordial black holes \cite{CKSY2010} or from accreting primordial black holes \cite{ROM2008,ABF2017,Am2017}. We note that there are examples of
  long lived composite particles
   in the standard cosmology. During the big bang nucleosynthesis, tritium
  (${}^3$H)
  and beryllium (${}^7$Be) are produced which decay to stable isotopes of helium (${}^3$He) and
  lithium (${}^7$Li) at redshifts of $z_{\rm 7Be}=3\times 10^4$ and $z_{\rm
    3H}=2.5\times 10^5$ respectively injecting energy into the CMB 
  \cite{ks2011}. Unfortunately the abundance of these elements in our
  Universe  is too low
  for the resulting spectral distortions to be observable.

 We use Planck \cite{Pl2018} $\Lambda$CDM
 cosmological parameters for all calculations and parameterize the
 injected energy, $\rho_{\rm X}$, as a fraction of CDM dark matter density,
 $f_{\rm X} = \rho_{\rm X}/\rho_{\rm CDM}$, where $\rho_{\rm CDM}$ is the
  $\Lambda$CDM (non-decaying) dark matter density. When we consider only
 $yim$-distortions, $f_{\rm X}$ and lifetime $\tau_X$ or the corresponding
  redshift $z_{X}$ are the only
 parameters that need to be considered. With the explicit evolution of
 electromagnetic cascades, the spectral distortions and constraints
 in addition become sensitive to the dark matter mass, $m_X$, as well as the
 the decay channel or the initial spectrum
 of photons, electrons and positrons injected in the decay.

  \section{Electromagnetic cascades in the expanding Universe.} In this work,
  we explicitly follow the
  electromagnetic cascade to calculate the $ntr$ part of the distortion while
  keeping track of energy that is lost to heat to calculate the
  $yim$-type contribution to the total CMB spectral distortion resulting
  from the injection of energetic particles. When using
  COBE-FIRAS data, previous
  authors have approximated the full distortion with a $y$-type or $\mu$-type
  distortion \cite{Hs1993,FRT20031,DKC2016}. We note that
  even in the \emph{all energy going to heat approximation} there is a
  small difference between the $y$- and $\mu$-distortion from the actual
  $yim$-distortion for
  $10^4\lesssim z \lesssim 2\times 10^5$ \cite{Ks2012b}. We
  evolve even the heat contribution to the spectral distortions with
  Kompaneets equation \cite{kom1956} getting correctly the intermediate or $i$-type
  part of the distortion and use the actual $yim$ and $ntr$-distortions for
  our COBE constraints.
  
At high redshifts($z\gtrsim
  2\times 10^5$), the Compton scattering process is very efficient and any initial
  distortion, including the $ntr$-type, thermalizes to a Bose-Einstein spectrum or $\mu$-distortion
  irrespective of energy and type of injected particles \cite{Ks2012b}. At
  $z\lesssim 2\times 10^5$, kinetic equilibrium is no longer possible and
  we must follow the full particle cascade to correctly calculate the final
  distortion.

  We divide the energy
 range of interest into energy bins for each particle (e.g. photons,
 electrons and positrons). The problem of electromagnetic cascade is then
 cast as a system of ordinary differential equations describing the flow of
 particles between different energy bins. This system can then be solved
 using the
 inductive approach
 \cite{Slatyer:2009yq,Kanzaki:2008qb,Kanzaki:2009hf}. Staring with an
 initial high energy particle, the electromagnetic cascade proceeds by
 sharing the total energy with more and more background particles, quickly
 multiplying the number of energetic particles
 (i.e. with energy $much$ greater than the background electrons and photons). 
 Since, the energy
cascade is one-way (i.e. higher to low energy), solution for a given energy
bin only depends on the cascade solution for the lower energy
bins. Therefore, starting with the lowest
energy bins, we can solve for populations of higher and higher energy
bins in the cascade. Details of our numerical codes implementing the above
method are described in detail in \citep{AK2018}. 

The inductive approach is an exact solution to the evolution
equations under the assumption that a particle in a particular energy bin
can only scatter to a lower energy bin.
This is true as long as the thermal energies of CMB photons and  background
electrons can be neglected at desired numerical accuracy. Once this is no
longer valid, which happens when the particles have become
non-relativistic, we evolve the full Kompaneets equation which takes into
account thermal distribution of background particles and scattering of
particles into higher energy bins. Our approximations are therefore driven
by desired numerical accuracy and not by ad hoc assumptions about the
physics. In this sense our calculations are exact. Previous constraints
were based on $yim$-distortions which are the solutions of Kompaneets
equation valid only in the non-relativistic limit. Therefore, the previous
constraints are inaccurate since they use non-relativistic solutions even
when the injected particles are relativistic. We should emphasize that we use Kompaneets equation only when the energies of photons and electrons are much smaller than the electron mass and the conditions for the validity of Kompaneets equation are satisfied. 

 In an ionized universe, the most important scattering process for low
 energy electrons($\lesssim$ keV) is Coulomb scattering and results in
 thermalization of the electron with most of its energy going to heat or
 $yim$-type distortions.  The dominant energy loss mechanism for a relativistic electron or
 positron is elastic scattering (inverse Compton) with the CMB.
  Once a positron becomes non-relativistic it annihilates with a
 background electron to give two  511 keV photons.   The injection of a
 positron is therefore, to a very good
 approximation, equivalent to an electron with same initial kinetic energy
 and two 511 keV photons. Since, the energy loss rates for electrons and positrons are extremely rapid compared to Hubble scale, they deposit their energy instantaneously \citep{Slatyer:2009yq,AK2018}. 

 For photons, the relevant scattering processes are Compton scattering,
 pair production and photon-photon elastic scattering. For most of the
 relevant energy range, the energy loss rate for photons is comparable to the
 Hubble rate. Therefore, we evolve the photon spectrum with background expansion
 taken into account.  The most important process which determines the
 shape of the spectral distortion is elastic or Compton scattering  of
 photons and electrons. Energetic photons produce energetic electrons and
 energetic electrons produce energetic photons by Compton scattering. The
 cycle repeats until the electrons start losing energy by Coulomb
 scattering to heat or loss of energy by photons in Compton scattering
 ($\propto \nu/m_{\rm e}$, where $\nu$ is the photon frequency and $m_{\rm
   e}$ is the electron mass)
 becomes inefficient, i.e. the spectral distortion is frozen.  Initial
 $\sim 10~{\rm keV}$ electrons  will just produce a
 spectral distortion with significant high energy Wien tail ($x=h\nu/(\kB T)\gtrsim 20$), where $h$ is Planck's
 constant and $\kB$ is the Boltzmann constant, and $T$ is CMB temperature) compared to $yim$-distortions. As we increase the energy of initial particles,
 they lose energy in smaller number of scatterings, giving a smaller
 amplitude  in the CMB band and more photons at higher and higher energies
 in the Wein tail. The highest energy photons in the distortion will however
 lose their energy to background electrons which will upscatter CMB and increase
 the amplitude of the distortion in the main CMB band. Thus the energy from the highest
 energy photons in the CMB spectral distortion cycles back to the main CMB
 band ($x\lesssim 20$). The amplitude of the distortions in
 the main CMB band therefore oscillates as we increase energy of injected
 particles or the dark matter mass.  In addition, due to the cyclical
 production of energetic electrons and photons, for very high energy ($\gtrsim$10 GeV) electron injection and photon injection become indistinguishable \citep{Slatyer:2009yq,AK2018}. 
 \begin{figure}
\centering
   \includegraphics[scale=1.0]{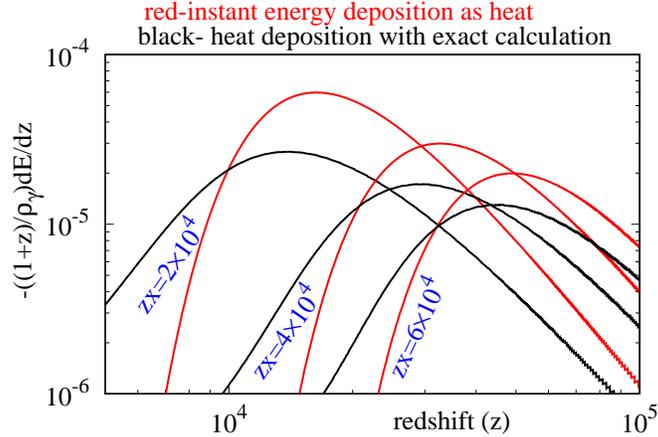}
   \caption{Ratio of energy density deposited as heat with respect to CMB
     energy density for varying $z_X$ and $f_X=0.0003$ for 20 GeV dark matter
     decay into monochromatic electron-positron pairs.}  
     \label{fig:heatbias}
  \end{figure}       
\begin{figure}[!tbp]
  \begin{subfigure}[b]{0.4\textwidth}
    \includegraphics[scale=1.0]{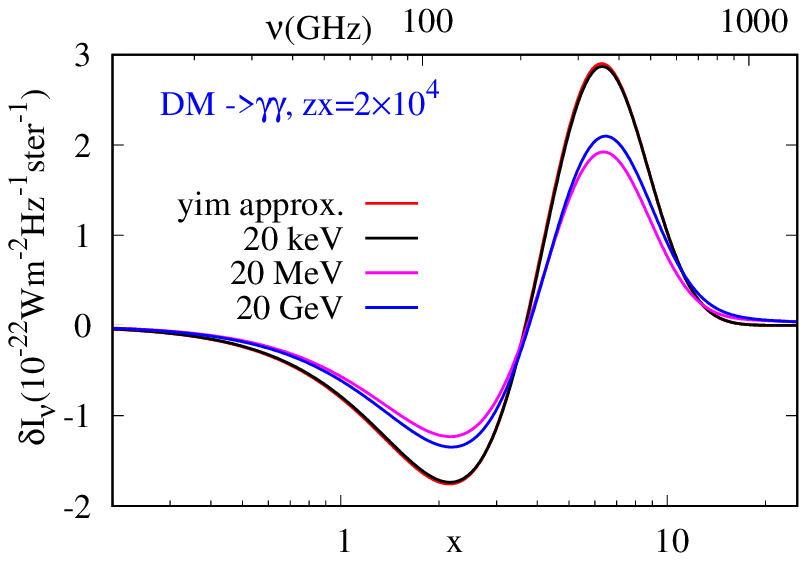}
  \end{subfigure}\hspace{65 pt}
  \begin{subfigure}[b]{0.4\textwidth}
    \includegraphics[scale=1.0]{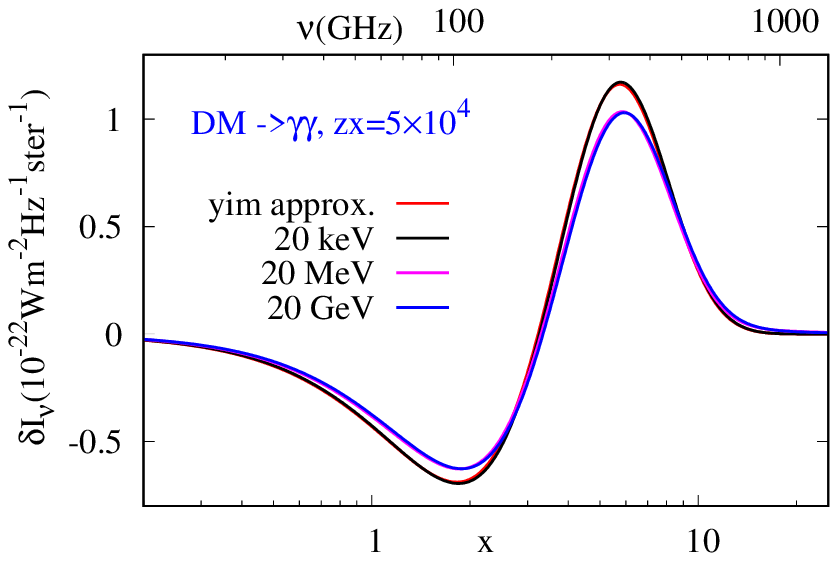}
    \end{subfigure}\\
    
    \begin{subfigure}[b]{0.4\textwidth}
    \includegraphics[scale=1.0]{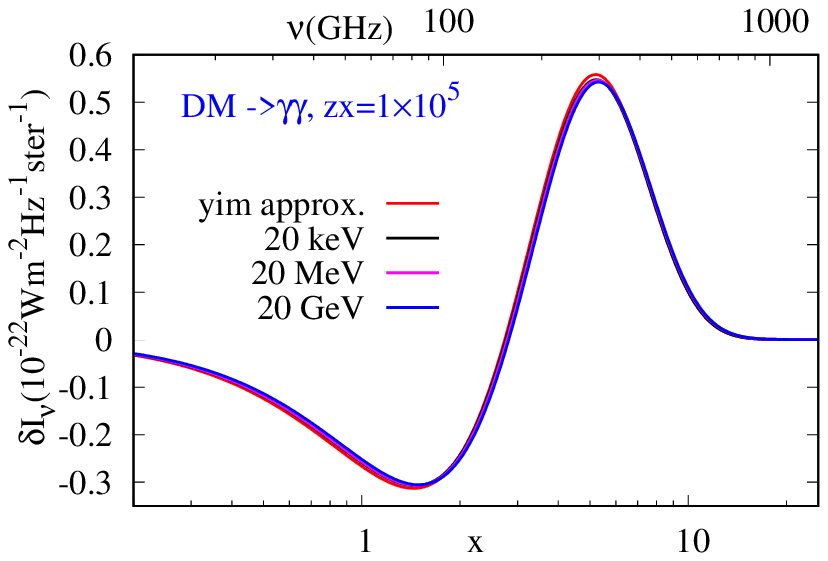}
  \end{subfigure}\hspace{65 pt}
  \begin{subfigure}[b]{0.4\textwidth}
    \includegraphics[scale=1.0]{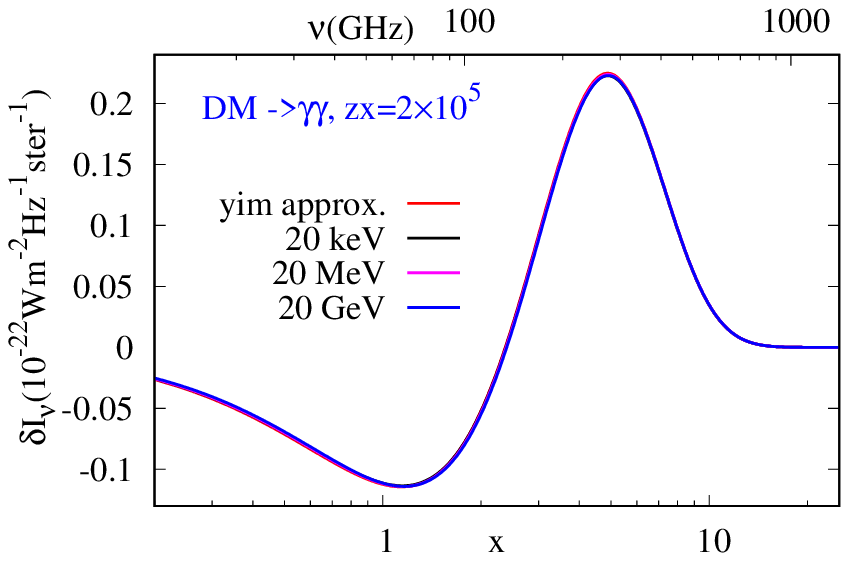}
   \end{subfigure}
   \caption{Spectral distortions  for dark matter decay into
     monochromatic photons for  different  $m_X$ and $z_X$
     and constant $f_X=0.0003$.}
  \label{fig:10kev10gevtherm}
\end{figure}
\begin{figure}[!tbp]
  \begin{subfigure}[b]{0.4\textwidth}
    \includegraphics[scale=1.0]{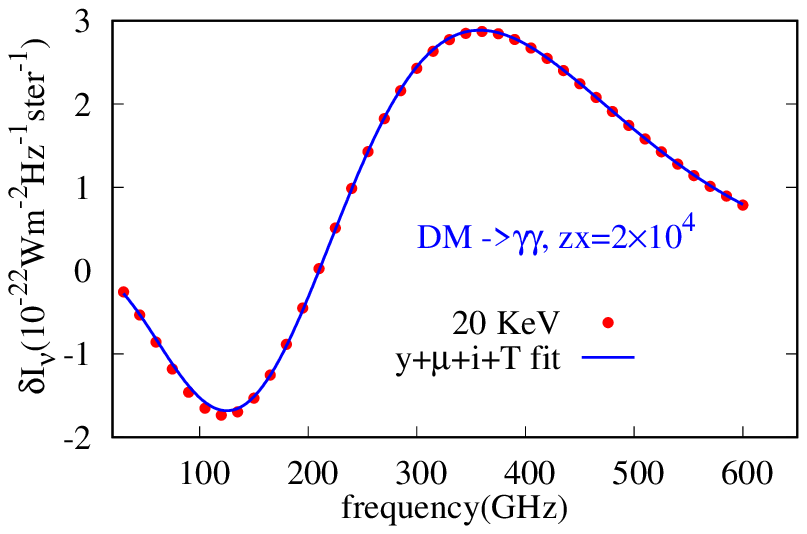}
  \end{subfigure}\hspace{65 pt}
  \begin{subfigure}[b]{0.4\textwidth}
    \includegraphics[scale=1.0]{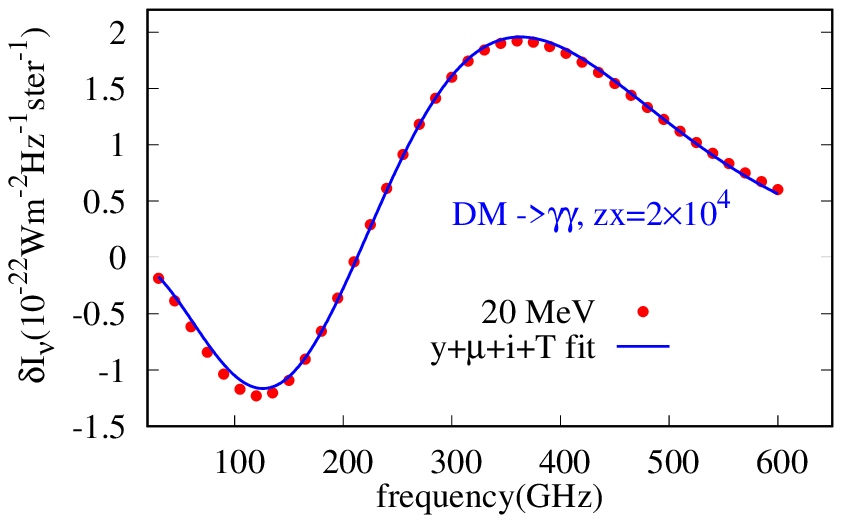}
    \end{subfigure}\\
    
    \begin{subfigure}[b]{0.4\textwidth}
    \includegraphics[scale=1.0]{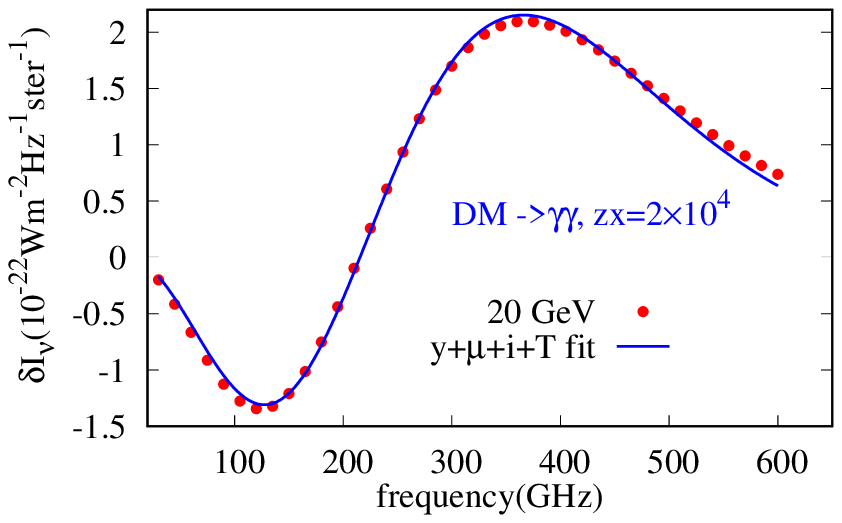}
  \end{subfigure}\hspace{65 pt}
  \begin{subfigure}[b]{0.4\textwidth}
    \includegraphics[scale=1.0]{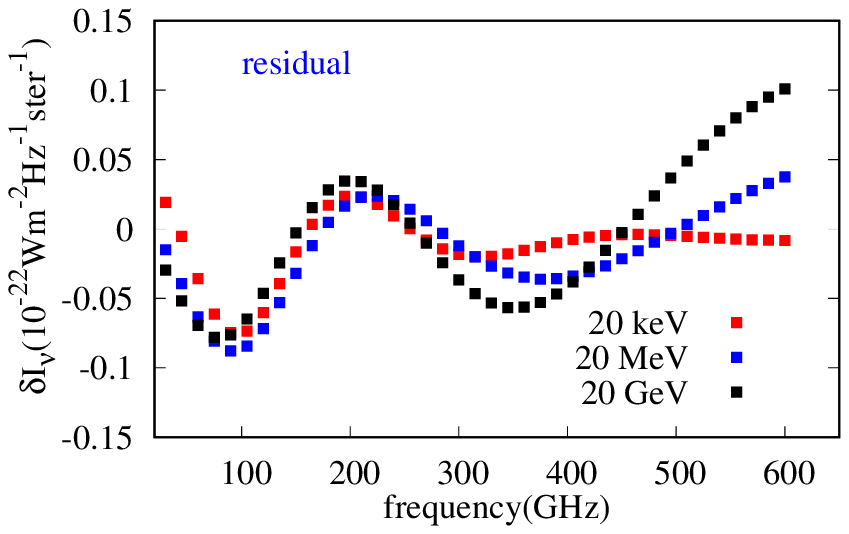}
   \end{subfigure}
   \caption{Best thermal distortion fit to actual distortion spectral and corresponding residual for decaying dark matter of lifetime $z_X$=20000. }
  \label{fig:residual}
\end{figure}
\begin{figure}[!tbp]
  \begin{subfigure}[b]{0.4\textwidth}
    \includegraphics[scale=1.0]{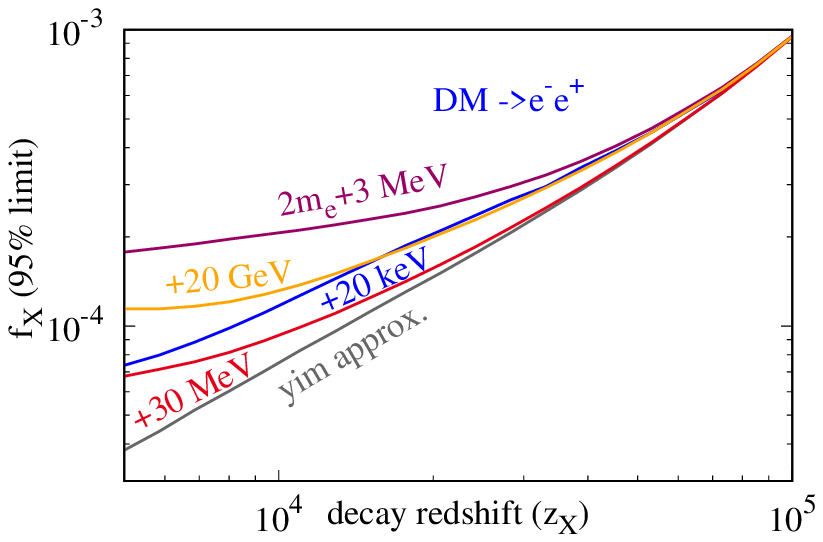}
  \end{subfigure}\hspace{65 pt}
  \begin{subfigure}[b]{0.4\textwidth}
    \includegraphics[scale=1.0]{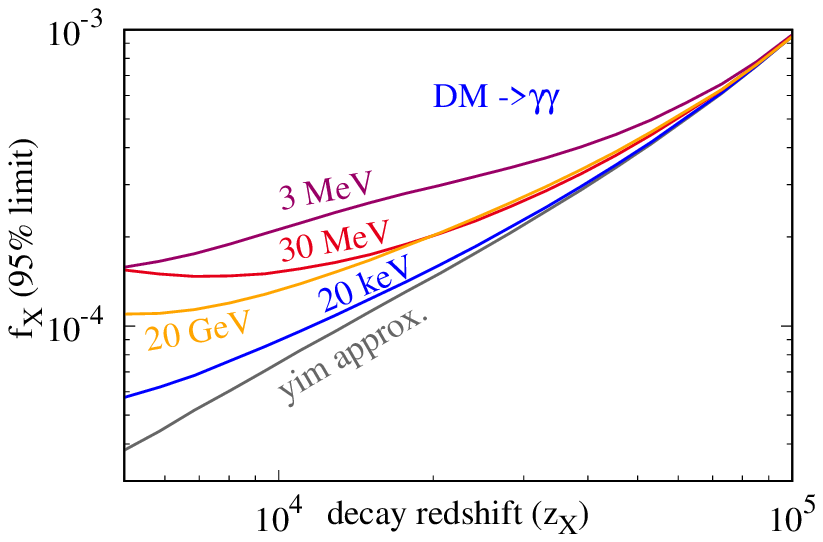}
    \end{subfigure}
    \caption{$95\%$ COBE-FIRAS upper limits on $f_X$ for different $m_X$ as
      a function of decay redshift $z_X$.
  Also shown is the constraint in the $yim$-approximation.}
  \label{fig:dmdecayconstraint}
\end{figure}
  \begin{figure}[!tbp]
  \begin{subfigure}[b]{0.4\textwidth}
    \includegraphics[scale=0.95]{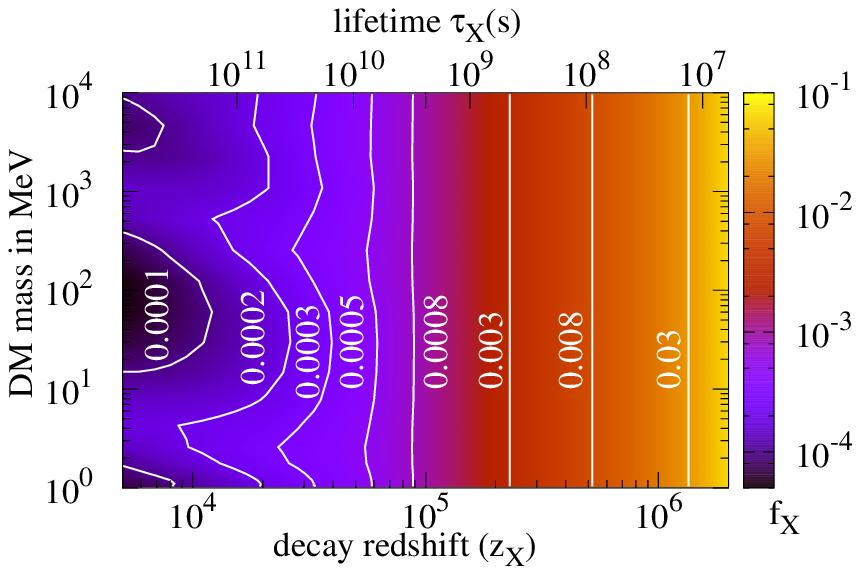}

  \end{subfigure}\hspace{65 pt}
  \begin{subfigure}[b]{0.4\textwidth}
    \includegraphics[scale=0.95]{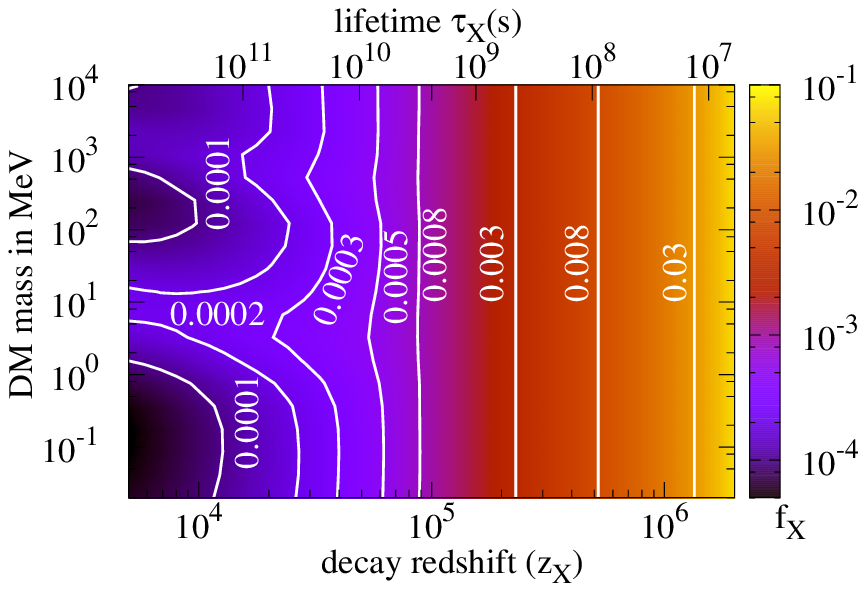}
    \end{subfigure}\\
   \caption{$95\%$ COBE-FIRAS limits on $f_X$ for decay into
     electron-positron pairs (left) and photons (right).}
  \label{fig:dmdecayconstraint1}
\end{figure}

\section{Spectral distortions from dark matter decay into photons and
  electron-positron pairs.} 
   We can write the energy density ($E$) injection rate for particle decay  as
\begin{equation}
 \frac{dE}{dt}=\frac{f_X \rho_{\rm DM}}{\tau_X} \exp(-t /\tau_X)
\end{equation} 
  where $\tau_X$ is the particle decay lifetime and $z_X$ is the
  redshift at proper time $t=\tau_X$, $\rho_{\rm DM}(z)=(1+z)^3\rho_{\rm DM}(z=0)$ is the non-decaying
  dark matter energy density at redshift
  $z$ and $f_X$ is ratio of the initial energy density of decaying dark matter
to that of the non-decaying component. 

We plot the energy density deposited as heat for  20 GeV dark matter decaying into 10 GeV electron-positron pairs per logarithmic redshift
interval as a fraction of the CMB energy
density ($\rho_{\gamma}$)  with full evolution of
particle cascades in Fig. \ref{fig:heatbias} and compare it with the
``$yim$-approximation'' i.e. instantaneous deposition of all energy as heat for
$f_X=0.0003$. There is a clear time delay between energy injection and
deposition as heat in the exact calculation and there is a bias in the temporal
information also in the $yim$-approximation.

In Fig. \ref{fig:10kev10gevtherm}, we plot the spectral distortions 
for  dark matter decaying to two
monochromatic photons and compare our exact  calculations with the
 spectral distortion in the $yim$-approximation, the latter being
 insensitive to the  mass of dark matter. In our new   calculation we find that the amplitude
 of the distortion is very sensitive to the mass of the dark matter, especially at low redshifts. The
 difference in shape is most evident in the high frequency tail of the
 spectral distortion. As we increase the redshift, the sensitivity to the
 initial dark matter mass decreases and the spectral distortion becomes
 closer to the $yim$-approximation, the two becoming indistinguishable and
 close to a Bose-Einstein spectrum ($\mu$-distortion) at $z_X\gtrsim 2\times 10^5$.

To see the difference between our new calculations and the
 $yim$-distortions, we fit the $yim$-distortions to our solutions for a PIXIE like experiment \cite{Pixie2011}. We fit the
   sum of $y$-type, $\mu$-type, $i$-type at redshift 20000, and a temperature
   shift (see Appendix C of \cite{AK2018})  with the amplitude of the four spectra as free parameters. For the fit we  sample the distortions at equally spaced PIXIE channels from 30
   GHz to 600 GHz with 15 GHz spacing between the channels. The results of
   such a fit are shown in Fig.\,\ref{fig:residual} for some of the decay channels shown in Fig. \ref{fig:10kev10gevtherm} for
   decaying dark matter with decay redshift $z_X=2\times 10^4$. 
    We also show the residuals which are the  difference between our calculation and the best fit $yim$-distortions. The
    residuals are
    especially large at high frequencies as can be already seen in Fig. \ref{fig:10kev10gevtherm}. The presence of high frequency channels and efficient
    removal of high frequency foregrounds will therefore be important for
    detecting the non-thermal distortions in the future experiments. 

As long as we do not have a detection but only upper
limits, we can get conservative limits by considering only one type of
distortion at a time. This is for example also done by the COBE-FIRAS
\cite{F1996}  team when they give limits on $y$ and $\mu$-type distortions
separately. However, once we have a positive detection of distortion there
will be  degeneracy between post recombination and pre recombination
distortions. The only way such a degeneracy can be broken is if we can
detect  the non-thermal part of the pre-recombination distortions. This
approach will however not work for the $yim$-distortions, e.g.  from Silk
damping \cite{Cks2012,Ksc20121}.  Another way would be using 21cm
observations to learn about the reionization epoch and subtract the
reionization contribution. Pre-recombination distortions would also modify
the cosmological recombination spectrum and can in principle be used to
constrain pre-recombinational energy injection \cite{Chluba:2008aw}.
    
We use COBE-FIRAS data \cite{F1996}, publicly available on the NASA LAMBDA
website\footnote{\url{https://lambda.gsfc.nasa.gov/product/cobe/firas_products.cfm}}, to constrain the fraction of dark matter that can decay as a
function of decay redshift $z_X$. We use the same procedure that was used by
\cite{F1996} for the $\mu$-type distortion and simultaneously fit the spectral distortion from decay, a
temperature shift and the galactic spectrum given in Table 4 of 
\cite{F1996} to the COBE-FIRAS monopole residuals. We do not find
any positive detection and plot $95\%$ upper limits for $f_X$ in
Fig. \ref{fig:dmdecayconstraint} for dark matter decaying to monochromatic
photons and electron-positron pairs.  We also show constraints in the
$yim$-approximation.   

If we assume that all energy goes into the spectral distortion with fixed shape, e.g. a
$\mu$-type distortion, then the amplitude of the distortion  will just be proportional to
$f_X\rho_{DM}/\rho_{\gamma} \propto f_X/(1+z_X)$. At higher redshifts the
decay of same amount of dark matter will give a smaller distortion because
the energy density of CMB is  higher by  a factor
of $(1+z_X)$. For a given sensitivity of the experiment, the constraints on
$f_X$ become weaker with redshift, $\propto (1+z_X)$, with some correction because the decay does not
happen instantaneously. This is what we see in the curve labeled $yim$-approximation. Since we take into account that the distortion at $z\lesssim
2\times 10^5$ is not exactly  $\mu$ but $i$-type and becomes  $y$-type at
$z\sim 10^4$, there is small departure from this linear relation.  For our exact calculation, explicitly evolving the
electromagnetic cascade, there is
dramatic 
departure from the $yim$-approximation, with the difference of a
factor of 4.7 for electron-positron channel for $m_X\approx 4~{\rm MeV}$ and a factor of 4.1 for the photon
channel for $m_X\approx 3~{\rm MeV}$ at  $z_X\approx 5000$ or
dark matter lifetime $\tau_X = 8\times 10^{11}{\rm s}$.  As we go to
higher redshifts, the constraints become closer to the $yim$-approximation
as Compton scattering become more efficient in reprocessing the energy
trying to establish a Bose-Einstein spectrum. The constraints in particular
are very sensitive to the  mass of the decaying particle. 

We show our new constraints from full evolution of electromagnetic
cascades in the $m_X-z_X,\tau_X$ plane in Fig. \ref{fig:dmdecayconstraint1}
for decay  to  photons and electron-positron pairs. The colour scale
represents the $95\%$ upper limits on $f_X$ from COBE-FIRAS data. In the
$yim$-approximation there will be no sensitivity to $m_X$ and the
isocontours of $f_X$ would all be vertical lines. This is what we see at
high redshifts, where $yim$-approximation holds. At lower redshifts the
spectral distortions and hence the constraints become sensitive to
$m_X$. The cyclic nature of electromagnetic cascades, as discussed above, is reflected in
oscillation in the isocontours in the $m_X$ direction. 
 At $z\approx 2\times 10^6$ the photon creation processes,
double Compton scattering and bremsstrahlung, become important suppressing
the spectral distortions \cite{Sz19701,dd1982} and weakening the constraints. We have taken this
suppression into account
using the  blackbody visibility function from \cite{ks2012}. 

 We should also mention that there are existing constraints from
big bang nucleosynthesis (BBN) with explicit evolution of  electromagnetic
cascades  \cite{kkm2004,kkt2018,ps2015,hsw2018,fmw2019} for decaying dark
matter with the same lifetimes as are considered in this paper. We refer
reader to Fig. 9 of \cite{fmw2019} for a comparision of constraints
obtained from spectral distortions and BBN. For photon injections with
energy close to photo-dissociation threshold of deuterium ($\lesssim 10$
MeV), the  constraints
from spectral distortions are stronger compared to BBN.  This is because these photons
immediately scatter with the background electrons dividing their energy
into  lower energy
photons below the deuterium dissociation threshold of 2.2 MeV and are
therefore unable to dissociate deuterium nuclei. As we increase the  energy
of the injected photons, more and more photons will be able to dissociate
 deuterium and, for high enough energy, helium nuclei. Hence constraints
 from BBN are stronger for photon injection with energies  $\gtrsim $10
 MeV. For electron-positron pair injection in the energy range  10 MeV-100
 MeV, the low energy photons produced by inverse Compton scattering are
 below the deuterium dissociation threshold, though there are some high
 energy photons produced by the final state radiation. For higher energies,
 we expect the constraints from $e^-e^+$ to be stronger than spectral
 distortion constraints and follow similar patterns as for the photon
 injection (e.g. see Fig. 13 of \cite{kkt2018}). Since, with our detailed calculations, constraints from spectral distortions are relaxed, for most of the energy range, constraints from BBN are still stronger compared to spectral distortion with present data.  Future experiments will improve the spectral distortion constraints by many orders of magnitude and accurate calculations such  as ours are essential to make the science case and to arrive at the minimum improvement in sensitivity that will be necessary to beat the BBN constraints.

\section{Conclusions}
  We have derived new upper limits on dark matter decay in the early Universe
from COBE-FIRAS measurements of the CMB monopole spectrum.
We explicitly evolve the electromagnetic cascades resulting from dark
matter decay into  photons and electron-positron pairs, calculate the
resulting spectral distortions in the CMB band and use these to constrain
the dark matter decay into these two channels. Previous COBE constraints have assumed that all
energy from decay goes
into heat and gives $y$-type and $\mu$-type distortions and were therefore blind to the decay channel. We show that these
approximations fail at low redshifts, $z_X\lesssim 2\times 10^5$ or dark
matter lifetimes $\tau_X \gtrsim 6\times 10^8~{\rm s}$. In addition, the
spectral distortions, and hence the constraints, are sensitive to the dark
matter mass. Our results show that the
decay channels are important for spectral distortions, just as they are
important for the CMB recombination history for larger lifetimes \cite{ASS1998,Chen:2003gz,sw2017,Pls2017}.
 This work motivates a more comprehensive study, in the future, taking into account different decay
channels motivated by different particle physics models in specific
cosmological scenarios such as in \cite{DKC2016}. Future
experiment like Primordial Inflation Explorer (PIXIE) \cite{Pixie2011} would
have a sensitivity of 3-4 orders of magnitude better than COBE-FIRAS \cite{fm2002} and
may discover distortions from new physics in the early Universe. Accurate
calculations of spectral distortions by explicitly evolving the
electromagnetic cascades would be crucial in
interpretation of such a future detection.

\section*{Acknowledgments:} This work was supported by Science and
Engineering Research Board, Department of Science and Technology, Govt. of
India grant  no. ECR/2015/000078. This work was also supported by Max
Planck Partner Group between Tata Institute of Fundamental Research, Mumbai and
Max-Planck-Institut f\"ur Astrophysik, Garching funded by Max-Planck-Gesellschaft.
\bibliographystyle{unsrtads}
\bibliography{spec-const2} 
\end{document}